\documentclass[12pt, doublespace]{article}

\usepackage{amsmath}
\usepackage{amssymb}

\newtheorem{definition}{Definition}
\newtheorem{theorem}{Theorem}

\newtheorem{proposition}{Proposition}
\newtheorem{corollary}{Corollary}

\newtheorem{lemma}{Lemma}
\def\cqd{\hfill \rule{2.25mm}{2.25mm}\vspace{10pt}}

\begin{document}

\title{Axiomatic structure of $k$-additive capacities\footnote{This paper is an
extended and revised version of \cite{migr00}, presented at IPMU2000 Conference.}} \author{P. Miranda\footnote{Corresponding author: P. Miranda. Dpt. of Statistics and O. R. and D.M. University of Oviedo. Address: c/ Calvo Sotelo s/n. 33007 Oviedo (Spain). Tel: (00) 34 985102955. e-mail: pmm@pinon.ccu.uniovi.es}\\University of Oviedo, Spain \and M. Grabisch\\ Universit\'e Paris I --
Panth\'eon-Sorbonne; LIP6 \and P. Gil\\ University of Oviedo, Spain} \maketitle
\begin{abstract}
In this paper we deal with the problem of axiomatizing the preference relations
modelled through Choquet integral with respect to a $k$-additive capacity, i.e.
whose M\"obius transform vanishes for subsets of more than $k$ elements. Thus,
$k$-additive capacities range from probability measures ($k=1$) to general capacities ($k=n$). The axiomatization is done in
several steps, starting from symmetric 2-additive capacities, a case related to the Gini index, and finishing with general $k$-additive capacities.
We put an emphasis on 2-additive capacities. Our axiomatization is done in the
framework of social welfare, and complete previous results of Weymark, Gilboa
and Ben Porath, and Gajdos. 

{\bf Classification code: C43, D31, D63.}

\end{abstract}

\newpage 

\section{Introduction}
Since the pioneering works of Schmeidler \cite{sch86, sch89} and
of many followers, capacities (otherwise called non-additive
measures by Denneberg \cite{den94} or fuzzy measures by Sugeno
\cite{sug74}) and the Choquet integral \cite{cho53} have become an
important tool in decision making, as a generalization of
probability measures and expected value. The main concept
underlying models based on capacities is the one of
\emph{comonotonic} functions (or acts, alternatives), i.e.
functions $f,g$ such that $(f(x)-f(x'))(g(x)-g(x'))\geq 0$, for
any $x,x'$. For such functions, the Choquet integral becomes a
linear operator, and many axiomatic characterizations of decision
making models based on capacities contain axioms restricted to
comonotonic acts, hence weakening corresponding axioms of
probability-based models. This is the case for the axiomatic
proposed by Schmeidler \cite{sch86}, where the famous independence
axiom is restricted to comonotonic acts (see also e.g. Chateauneuf
\cite{cha94}, Wakker \cite{wak89}).

Capacities have also been widely used in decision under multiple
criteria. Let $X$ be a set of $n$ criteria. A capacity
$\mu:\mathcal{P}(X)\longrightarrow [0,1]$ is defined on $X$, and
for any subset of criteria $A\subseteq X$, roughly speaking
$\mu(A)$ represents the importance of the set $A$ of criteria for
the decision making problem into consideration. Avoiding
intricacies, an alternative can be viewed as a function
$f:X\longrightarrow \mathbb{R}$, where $f(x)$ stands for the score
of alternative $f$ w.r.t. criterion $x\in X$. Then the Choquet
integral of $f$ represents the overall score of alternative $f,$
taking into account the importance of criteria, modelled by $\mu$.
The axiomatic justification and building of such a model has been
given in \cite{grlava03, lagr03}. The major interest of such
models is that they enable a proper representation of interaction
between criteria, a topic of high importance in applications (see
e.g. \cite{grdulipe02}).

This flexibility of capacity-based models has to be however paid
by an exponential complexity, an important drawback in practice,
since a capacity on a set of $n$ elements needs $2^n$ real values
to be defined. A solution to this problem is to work with some
sub-families of capacities, requiring less coefficients to be
defined. Among them, the sub-families of \emph{$k$-additive
measures} (which may be called also \emph{$k$-additive
capacities}), proposed by Grabisch \cite{gra96c, gra97c} have the
interest to be nested, starting with classical (additive) measures
($k=1$), and ending with non-additive measures in their full
generality ($k=n$). The idea in fact stems from pseudo-Boolean
functions $f:\{0,1\}^n\longrightarrow \mathbb{R}$, which are
another view of set functions \cite{haho92}. It is known that they
can be expressed under a polynomial form of degree $n$, involving
at most $2^n$ terms. Hence, a $k$-additive measure $\mu$ is simply
a non-additive measure whose corresponding pseudo-Boolean function
has a polynomial development of degree at most $k$. Interestingly
enough, this amounts to say that the M\"obius transform of $\mu$
vanishes for subsets of more than $k$ elements, and also that
there is no interaction among subsets of more than $k$ elements
\cite{gra97c}. $k$-additive measures (and especially 2-additive
measures, since being more general than additive ones, while
remaining simple) have been successfully used in multicriteria
decision making (see e.g. \cite{grdulipe02}).

One question remains however open: what about the axiomatic
characterization of $k$-additive measures? In decision making, and
especially in decision under uncertainty or risk, the main issue
is to know the properties of preference structures underlying a
given mathematical model. $k$-additive measures have proven their
usefulness in practice, but it remains to know what precisely they
imply or allow for the representation of the preference of the
decision maker. Our aim in this paper is to fill this gap. We will
see that our axiomatic has its roots in social welfare theory, and
is related to previous works by Weymark \cite{wey81}, and Ben
Porath and Gilboa \cite{pogi94}.

The paper is organized as follows: We introduce some basic
concepts in Section 2. Then, in Section 3 we deal with the problem
of characterizing the preference relation when $\mu $ is a
symmetric capacity; from this starting result, we characterize
2-additive symmetric capacities (Section 4) and $k$-additive
symmetric capacities (Section 5).

We deal with the same problem in Sections 6, 7 and 8 removing the
symmetry condition. We will prove that results from Section 3 can
be straightforwardly applied to the general case, just removing
the symmetry axiom. However, for the 2-additive and the
$k$-additive cases (Sections 7 and 8), we will be forced to modify
the axioms obtained in Sections 4 and 5, respectively.

Finally, in Section 9 we give some conclusions.

\section{Basic concepts and notations}

In this paper, the universal set $X=\{ 1,..., n\} $ denotes a
finite set of $n$ elements (states of nature, criteria,
individuals, etc). The set of all subsets of $X$ is denoted ${\cal
P}(X)$. Subsets of $X$ are denoted by $A, B, \ldots .$ We will
sometimes write $i_1 \cdots i_k$ instead of $\{ i_1, \ldots ,
i_k\} $ in order to avoid heavy notation, specially with
singletons and subsets of two elements. We start by recalling some
definitions.

\begin{definition}\label{mesflo}
A {\bf capacity} \cite{cho53} or {\bf non-additive measure}
\cite{den94} or {\bf fuzzy measure} \cite{sug74} over $X$ is a
mapping $\mu :{\cal P}(X) \rightarrow [0,1]$ such that
\begin{itemize}
\item $\mu (\emptyset )=0,\, \mu (X)=1$ (boundary conditions).
\item $\forall A, B\in {\cal P}(X)$ such that $A\subseteq B$, we
have $\mu (A)\leq \mu (B)$ (monotonicity).
\end{itemize}
\end{definition}

As a special case of capacities, we have symmetric capacities.

\begin{definition}\label{symmea}
A capacity $\mu $ is said to be {\bf symmetric} if for any $A,
B\in {\cal P}(X)$ such that $ |A|=|B|,$ we have $\mu (A)=\mu (B).$
\end{definition}

The M\"obius transform is an invertible linear transform of set
functions, and is a fundamental notion in capacity theory
\cite{chja89}.

\begin{definition}\cite{rot64}\label{defmobinv}
Let $\mu:\mathcal{P}(X)\rightarrow \mathbb{R} $ be a set function
on $X$. The {\bf M\"obius transform (or inverse)} of $\mu $ is
defined by
$$
m(A):=\sum_{B\subseteq A} (-1)^{|A\setminus B|} \mu (B), \,
\forall A\subseteq X.
$$ The M\"obius transform being given, $\mu$ is recovered by the
Zeta transform
\begin{equation}
\label{eq:mob}
\mu(A)=\sum_{B\subseteq A}m(B).
\end{equation}
\end{definition}

It is trivial to see that a capacity $\mu $ is symmetric if and
only if for any $A, B\in {\cal P}(X)$ such that $ |A|=|B|,$ we
have $m (A)=m (B).$

The monotonicity constraints in terms of $m$ are given by:

\begin{proposition}\cite{chja89}\label{monconmob}
A set of $2^n$ coefficients $m(A), A\subseteq X$ corresponds to
the M\"obius representation of a capacity if and only if
\begin{enumerate}
\renewcommand{\labelenumi}{(\roman{enumi})}
\item $\displaystyle{ m(\emptyset )=0,\, \sum_{A\subseteq X}
m(A)=1,}$ \item $\displaystyle{\sum_{i\in B\subseteq A} m(B)\geq
0}$, for all $A\subseteq X$, for all $i\in A$.
\end{enumerate}
\end{proposition}

Another well-known example of non-additive measures are belief
functions:

\begin{definition}\cite{dem67, sha76}\label{bel}
A {\bf belief function} is a capacity $Bel: {\cal P}(X)\rightarrow
[0,1]$ satisfying for any family $A_1, ..., A_k\subseteq X$ the
following property:
\[
\mathrm{Bel}(\bigcup_{i=1}^k A_i) \geq \sum_{\emptyset\neq
I\subset\{1,\ldots,k\}}(-1)^{|I|+1}\mathrm{Bel}(\bigcap_{i\in
I}A_i),
\]
for any $k\geq 2.$
\end{definition}

Related to belief functions, the following result can be proved:

\begin{proposition}\cite{sha76}\label{conbel}
$\mu $ is a belief function if and only if its corresponding
M\"obius inverse is non-negative.
\end{proposition}

\begin{definition}\cite{gra96c}\label{kad}
A capacity $\mu $ is said to be {\bf $k$-order additive} or {\bf
$k$-additive} for short for some $k\in\{1,\ldots,n\}$ if its
M\"obius transform vanishes for any $A\subseteq X$ such that
$|A|>k,$ and there exists at least one subset $A$ of exactly $k$
elements such that $m(A)\not= 0.$
\end{definition}

We call any function $f:X\rightarrow \mathbb{R}$ an \emph{act} or
\emph{alternative}, and we denote $\mathcal{F}:=\mathbb{R}^X$ the
set of all acts on $X$, while $\mathcal{F}_M$ is the set of non decreasing acts. For convenience, we will often
denote $f(i)$ by $f_i$, and identify $f$ with the vector
$(f_1,\ldots,f_n)$ of its values. Constant acts $f_i=\alpha$ for
all $i\in X$ are denoted by $\alpha$ if no confusion occurs.

\begin{definition}\cite{halipo52}\label{com}
A pair of acts $f, g$ on $\mathcal{F}$ is said to be {\bf
comonotone} if and only if
$$ (f(i)-f(j))(g(i)-g(j))\geq 0,\quad \forall i,j\in X.$$
\end{definition}

\begin{definition}\cite{cho53}\label{cho}
Let $\mu $ be a capacity over $X$ and an act $f\in\mathcal{F}$.
The {\bf Choquet integral} of $f$ with respect to $\mu $ is
defined by $$ {\cal C}_{\mu }(f):=\int_0^{\infty }\mu (f\geq
\alpha )d\alpha + \int_{-\infty }^0 (\mu (f\geq \alpha )-1)d\alpha
,$$ which, since  $X$ is finite, reduces to $$ {\cal C}_{\mu
}(f)=\sum_{i=1}^n (f_{(i)}-f_{(i-1)})\mu (B_i),$$ where
$_{(\cdot)}$ stands for a permutation on $X$ such that $
0=:f_{(0)}\leq f_{(1)}\leq \cdots\leq f_{(n)}$, and $ B_i:=\{
(i),\ldots,(n)\}. $
\end{definition}

Choquet integral in terms of M\"obius transform is given by:

\begin{proposition}\cite{wal91}\label{chomob}
Let $\mu $ be a capacity and $m$ its M\"obius transform. Then, the
Choquet integral of $f\in\mathcal{F}$ with respect to $\mu $ in
terms of $m$ is expressed by:
$$ {\cal C}_{\mu } (f)=\sum_{A\subseteq X} m(A)
\bigwedge_{i\in A} f_i.$$
\end{proposition}

\begin{definition}\cite{yag88}\label{owa}
An {\bf ordered weighted averaging operator (OWA)} is an operator
on $\mathcal{F}$ defined by
$$ \mathrm{OWA}_w(f):=\sum_{i=1}^nw_if_{(i)},$$ where
$ w=(w_1, \ldots,w_n)\in [0,1]^n$ is such that
$\displaystyle{\sum_{i=1}^nw_i=1},$ (called a weight vector), and
$f_{(i)}$ is defined the same way as for Choquet integral.
\end{definition}

OWA operators and symmetric capacities are related through the
following result:
\begin{proposition}\label{owasym}\cite{gra95b, gra96b, musu93}
Let $\mu $ be a capacity on $X$. Then, the following statements
are equivalent:
\begin{enumerate}
\item
There exists a weight vector $w$ such that ${\cal C}_{\mu
}(f)=\mathrm{OWA}_w(f)$, for any $f\in\mathcal{F}$.
\item
$\mu $ is a symmetric capacity.
\end{enumerate}
\end{proposition}

Finally, we consider a preference relation $\succeq $ on
$\mathcal{F}\times \mathcal{F}$, assumed to be reflexive, transitive and
complete. As usual, the symmetric part of $\succeq$ is denoted
$\sim$, while $\succ$ denotes the asymmetric part. We say that
$V:\mathcal{F}\rightarrow \mathbb{R}$ is a
\textbf{representation} of $\succeq$ if for any pair of acts
$f,g\in\mathcal{F}$, we have
\[
f\succeq g\Leftrightarrow V(f)\geq V(g).
\]
Our goal in next sections will be to find a set of axioms over
$\succeq $ such that the Choquet integral w.r.t. a $k$-additive
capacity is a representation of $\succeq$.

\section{Characterization of OWA operators}

In this section we deal with the problem of characterizing the
preference relation induced by an OWA operator, i.e. the Choquet
integral with respect to a symmetric capacity (Proposition
\ref{owasym}). Let us consider an $\mathrm{OWA}_w$ operator, with
weight vector $w=(w_1,\ldots,w_n),$ and the preference relation
defined on ${\cal F}$ by $$ f\succeq g\Leftrightarrow
\sum_{i=1}^nw_if_{(i)}\geq \sum_{i=1}^nw_ig_{(i)}.$$

We introduce the following axioms, defined and interpreted in
\cite{wey81}, in the context of social welfare:

\begin{itemize}
\item {\bf A1.} Weak order: $\succeq $ is complete, reflexive and transitive.
\item {\bf A2.} Continuity: For every $f\in \mathcal{F}$, if
$\succ $ denotes the strict preference, the sets $\{ g\in
\mathcal{F}|g\succ f\} $ and $\{ g\in \mathcal{F}|g\prec f\} $
are open sets (in the topology of $\mathcal{F}$ induced by the
natural topology on $\mathbb{R}^n$). 
\item {\bf A3.} Symmetry:
For every $f, g\in \mathcal{F}$, if there is a permutation $\pi
$ on $X$ such that $f=\pi g$, then $f\sim g$. 
\item {\bf A4.} Weak independence of income source: For all  acts $f, g, h\in
\mathcal{F}_M,$ $f\succeq g\Leftrightarrow f+h\succeq g+h$.
\end{itemize}

With these axioms, Weymark proved:

\begin{theorem}\label{wey}\cite{wey81}
Let $\succeq $ be a preference relation over $\mathcal{F}\times
\mathcal{F}$. Then, $\succeq $ satisfies {\bf A1}, {\bf A2},
{\bf A3} and {\bf A4} if and only if $\exists \, {\cal
H}:\mathbb{R}^n\longrightarrow \mathbb{R}$ which is a
representation of $\succeq$, such that $${\cal H}(f)=\sum_{i=1}^n
w_if_{(i)}, $$ where $f_{(1)}\leq f_{(2)}\leq \cdots\leq f_{(n)}$,
and $w_i\in \mathbb{R} .$
\end{theorem}

Note that ${\cal H}$ is not an OWA operator, since $w$ is not
necessarily a weight vector.  Then, it suffices to add some axioms
to Weymark's system in a way such that they guarantee that the
operator obtained in Theorem \ref{wey} is an OWA operator.  For
the sake of simplicity, we will establish our result in several
propositions.

First, as a consequence of Proposition \ref{owasym}, the following
can be proved:

\begin{lemma}\label{axisym}
Let us suppose that we are given an operator defined for any
$f\in\mathcal{F}$ by $$\displaystyle {\cal
H}(f)=\sum_{j=1}^nw_jf_{(j)}.$$ Then, this operator is the Choquet
integral with respect to a symmetric capacity if and only if the
following conditions hold:

\begin{itemize}
\item
{\bf C1.} $w_i\geq 0,\, \forall i.$
\item
{\bf C2.} $\displaystyle{\sum_{i=1}^nw_i=1.}$
\end{itemize}
\end{lemma}

{\bf Proof:} Remark that {\bf C1} and {\bf C2} are the conditions
for $w$ to be
 a weight vector. Then, applying Proposition \ref{owasym},
it is clear that if we have a Choquet integral with respect to a
symmetric capacity, then {\bf C1} and {\bf C2} hold.

The reciprocal is also true. Just recall \cite{fomaro95} that the
capacity is defined as $$ \begin{array}{lcll}\mu (\{ i\} ) & = &
w_n, & \forall i  \\ \mu (\{ i,j\} ) & = & w_n+w_{n-1}, & \forall
i,j
\\ \vdots & \vdots & \vdots & \vdots \end{array} $$
whence the result. $\cqd $

Let us consider the following axiom:
\begin{itemize}
\item
{\bf A5.} Monotonicity: Given $f,g\in \mathcal{F}$, if $f_i\geq g_i,\, \forall i\Rightarrow
f\succeq g$.
\end{itemize}

Axiom {\bf A5} states that if an act $f$ is better or equal than
another act $g$ for all criteria, then this act must be considered
better or equal in our preference relation.

It is clear that the Choquet integral satisfies {\bf A5}. We
have the following proposition:

\begin{proposition}\label{equc1}
Given a preference relation over $\mathcal{F}\times
\mathcal{F}$, if {\bf A1}, {\bf A2}, {\bf A3} and {\bf A4} hold,
then {\bf A5} is equivalent to {\bf C1}.
\end{proposition}

{\bf Proof:} As {\bf A1}, {\bf A2}, {\bf A3}, {\bf A4} hold,
applying Theorem \ref{wey} our binary relation can be represented
by $$ {\cal H}(f)=\sum_{i=1}^nw_if_{(i)}.$$

$\Rightarrow )$ Let us suppose that {\bf A5} holds and that there
exists $i$ such that $w_i<0$. Let $f, g$ be defined by $g_j=0,\,
\forall j\leq i,\, g_j=1,\, \forall j>i,\, f_j=0,\, \forall j<i,\,
f_j=1,\, \forall j\geq i.$ Then, by {\bf A5}, $f\succeq g$.

On the other hand, if $w_i<0$, we have $$ {\cal H}(f)=\sum_{j=1}^n
w_jf_{(j)}=\sum_{j=1}^n w_jg_{(j)}+w_i <\sum_{j=1}^n w_jg_{(j)}={\cal
H}(g)\Rightarrow f\prec g,$$ a contradiction. Thus, $w_i\geq 0$.

$\Leftarrow ) $ Suppose on the other hand $w_i\geq 0.$ If
$\displaystyle f_i\geq g_i,\, \forall i\Rightarrow \sum_{j=1}^n
f_{(j)}w_j\geq \sum_{j=1}^n g_{(j)}w_j\Rightarrow f\succeq g,$ and
hence {\bf A5} holds. $ \cqd $

Let us now consider the following axiom:

\begin{itemize}
\item {\bf A6.} Non-triviality: There exist $f, g\in
\mathcal{F}$ such that $f\succ g$.
\end{itemize}

Axiom {\bf A6} is just needed in order to avoid the trivial
relation, i.e. the preference relation in which all alternatives
are considered equally good by the decision maker.

Then, the following proposition holds:

\begin{proposition}
Consider a preference relation over $\mathcal{F}\times
\mathcal{F}$. If {\bf A1}, {\bf A2}, {\bf A3}, {\bf A4}, {\bf
A5} hold, then {\bf A6} is equivalent to
$\displaystyle{\sum_{i=1}^n w_i>0}$.
\end{proposition}

{\bf Proof:} By Proposition \ref{equc1} we know that our binary
relation can be represented by $$ {\cal
H}(f)=\sum_{i=1}^nw_if_{(i)},\, w_i\geq 0,\, \forall i.$$

$\Rightarrow )$ Now, if $w_i=0,\, \forall i\Rightarrow {\cal
H}(f)=0,\, \forall f\in \mathcal{F},$ and then $f\sim g,\,
\forall f, g \in \mathcal{F},$ contradicting our hypothesis
({\bf A6}).

$\Leftarrow )$ If $\displaystyle{\sum_{i=1}^n w_i>0\Rightarrow
{\cal H}(1)>0={\cal H}(0)\Rightarrow 1\succ 0.}\cqd $

Now, we can normalize, and hence $\displaystyle{\sum_{i=1}^n
w_i=1}$. Then, applying Lemma \ref{axisym} and Proposition
\ref{owasym} we obtain the following:

\begin{theorem}\label{carowa}
\cite{migr00} Let $\succeq $ be a binary relation on $\mathcal{F}\times
\mathcal{F}$. The following statements are equivalent:
\begin{enumerate}
\item $\succeq $ satisfies {\bf A1}, {\bf A2}, {\bf A3}, {\bf A4},
{\bf A5} and {\bf A6}. \item There is a unique symmetric capacity
$\mu $ such that $\succeq $ is represented by ${\cal C}_{\mu }.$
\end{enumerate}
\end{theorem}

\section{Characterization of 2-additive OWA}

In this section we deal with the problem of characterizing
preferences represented by the Choquet integral with respect to a
2-additive symmetric capacity. In order to do this, we are going
to use a result proved by Ben Porath and Gilboa in \cite{pogi94}.
Let us consider the following definition:

\begin{definition}
Let $f$ be an act. We say that $i$ {\bf $f$-precedes} $j$ if $f_i<
f_j$ and there is no $k\in X$ such that $f_i<f_k<f_j.$
\end{definition}

Ben Porath and Gilboa use the set of axioms of Weymark except {\bf
A4} which is changed into {\bf A4'}; {\bf A5} is changed into a
stronger version {\bf A5'}. Finally they also add some other
axioms. Specifically:

\begin{itemize}
\item {\bf A4'.} Order-preserving gift: For every $f, f', g, g'$
in $\mathcal{F}_M$, for every $i\in X$, if $f_j=f'_j$ and
$g_j=g'_j$ for every $j\not= i$ and $f'_i=f_i+t, g'_i=g_i+t$ for
some $t\in \mathbb{R}$, then $f\succeq g$ if and only if
$f'\succeq g'$.

\item {\bf A5'.} Strong monotonicity: For every $f, g\in {\cal
F},$ if $f_i\geq g_i$ for all $i\in X$, and there exists a $j$
such that $f_j>g_j$, then $f\succ g$.

\item {\bf A7.} Order-preserving transfer: For every $f, f', g,
g'$ in $\mathcal{F}$, and for all $i, j\in X$, if $i$ $f-, g-,
f'-$ and $ g'-$precedes $j$, if $f'_i=f_i+t, g'_i=g_i+t,
f'_j=f_j-t, g'_j=g_j-t$ for some $t>0$, and $f'_k=f_k, g'_k=g_k$
for all $k\not= i, j$, then $f\succeq g$ if and only if $f'\succeq
g'$.

\item {\bf A8.} Inequality aversion: For every $f, f'$ in
$\mathcal{F}_M$, for all $i\in X$, if $f'_i=f_i+t,
f'_{i+1}=f_{i+1}-t$ for some $t>0$, if $f'_j=f_j$ for all $j\not=
i, i+1$, then $f'\succ f$.
\end{itemize}

Ben Porath and Gilboa work in the field of social welfare; thus, they consider
$X$ as a set of individuals, and acts are considered as the
income profile distribution of a society. Then, they try to find a society
where individuals are all equally rich, i.e where inequalities are reduced by
sharing wealth. An explanation of the axioms can be found in
\cite{pogi94}.

They proved the following:

\begin{theorem}\label{gilpor}
\cite{pogi94} Let $\succeq $ be a binary relation on $\mathcal{F}\times \mathcal{F}$. The following
statements are equivalent:
\begin{enumerate}
\item
$\succeq $ satisfies {\bf A1}, {\bf A2}, {\bf A3}, {\bf A4'}, {\bf A5'},
{\bf A7} and {\bf A8}.
\item
There is a unique number $\delta ,0<\delta <1/(n-1)$, such that
$\succeq $ is represented by the following functional:
\begin{equation}\label{eqgilpor}
{\cal H}(f)=\sum_{i\in X}f_i-\delta \left[ \sum_{1\leq i<j\leq
n}|f_i-f_j|\right] .
\end{equation}
\end{enumerate}
\end{theorem}

Grabisch \cite{gra98b} has shown that Equation (\ref{eqgilpor}) is the Choquet
integral with respect to a 2-additive symmetric capacity. However, this result
does not cover all of them. The problem comes from the fact that for this
result, the coefficient multiplying $f_{(i)}$ is $1+(n-2i)\delta $, and thus
the weights of the OWA operator are decreasing. We will prove below that {\bf
A8} is equivalent to have a strictly decreasing order in the weights of the OWA
operator. As this is not always the case for capacities, we have to remove
it. Indeed, the functional in Theorem \ref{gilpor} gives more importance to the
smallest value (the poorest individual) than to the biggest value (the richest
one) and thus, it tries to penalize the differences among incomes in the sense
that an increment in the poorest individual from a gift of the richest one
should lead to a better (preferred) income distribution. Anyway, we can
conclude that 2-additivity is given by {\bf A7}, {\bf A8} or a mixture of these
two axioms.

We will use this set of axioms in order to characterize 2-additive
symmetric capacities. As in the previous section, we do it in
several propositions.

\begin{proposition}\label{equiweibis}
Let us consider the Choquet integral w.r.t. a symmetric capacity $\mu$, and its
corresponding OWA operator with weight vector $w$. The following propositions are equivalent.
\begin{itemize}
\item [(i)] $\mu$ is at most 2-additive.
\item [(ii)] the weight vector has equidistant components
\[
w_i-w_{i+1}=w_1-w_2,\quad \forall i=1,\ldots, n-1.
\]
\end{itemize}
\end{proposition}

{\bf Proof:} Since $\mu$ is symmetric, the Choquet integral can be written
under its OWA form  (Proposition \ref{owasym}):
\begin{equation}\label{2owaaux2}
{\cal
C}_{\mu }(f)=\sum_{i=1}^nw_if_{(i)}.
\end{equation}

$(i)\Rightarrow(ii)$ since $\mu$ is a at most 2-additive, the Choquet integral
w.r.t. the M\"obius transform of $\mu$ writes (Proposition \ref{chomob})
\[
{\cal C}_{\mu }(f)=\sum_{i=1}^nm((i))f_{(i)}+\sum_{i<j}m((i),(j))f_{(i)}.
\]
Applying symmetry we have $m(A)=m(B)$ whenever
$|A|=|B|$. Therefore,
\begin{equation}\label{2owaaux1}
{\cal C}_{\mu
}(f)=\sum_{i=1}^n[k_1+(n-i)k_2]f_{(i)},
\end{equation}
with $k_1=m((i)), k_2=m((i),(j))$.  Identifying Equations
(\ref{2owaaux1}) and (\ref{2owaaux2}), we conclude $w_i=k_1+(n-i)k_2$ and thus
the weights are equidistant.

$(ii)\Rightarrow(i)$ Let us define $k_1:=w_n, k_2:=w_{n-1}-w_n$.
Then, $w_n=k_1, w_{n-1}=k_1+k_2$ and due to the hypothesis of
equidistance, we have $w_i=k_1+(n-i)k_2$. Thus, $$
\sum_{i=1}^nw_if_{(i)}=\sum_{i=1}^n[k_1+(n-i)k_2]f_{(i)}, $$ and
this is the Choquet integral of the 2-additive symmetric capacity
defined by $m(i) = k_1,\, m(i, j) = k_2$ (a 1-additive
capacity if $k_2=0$). As this is the only symmetric capacity whose
corresponding Choquet integral has these weights, the result is
proved. $\cqd $

In the case of 1-additive symmetric capacities, all coefficients
are equal and thus $w_1-w_2=0$.

%

Consequently, in order to characterize the preference relations
over $\mathcal{F}\times \mathcal{F}$ that are represented
through the Choquet integral with respect to a 2-additive (at
most) symmetric capacity, it suffices to find an axiom leading to
equidistance of the weights. This is solved in next proposition.

\begin{proposition}\label{axi2add}
Let $\succeq $ be a preference relation over $\mathcal{F}\times
\mathcal{F}$. If {\bf A1}, {\bf A2}, {\bf A3}, {\bf A4}, {\bf A5}, {\bf A6}
hold, then {\bf A7} is equivalent to $$w_1-w_2=w_i-w_{i+1},\, \forall
i=1,\ldots, n-1.$$
\end{proposition}

{\bf Proof:} $\Rightarrow )$ This is Lemma 3.2.1 in \cite{pogi94}.

$\Leftarrow )$ Let us consider $f, g\in \mathcal{F}$ and suppose
that $i$ $f-, g-$ precedes $j$. Suppose that $f\succeq g$. Then,
if $f_i$ is the $i_f$-th smallest income in $f$, and $g_i$ is the
$i_g$-th in $g$, defining $f'$ and $g'$ as in {\bf A7} and assuming that $i$ $f'-, g'-$ precedes $j,$ we have $$
{\cal H}(f')= \sum_{j=1}^n w_jf_{(j)}+t[w_{i_f}-w_{(i-1)_f}].$$ $$
{\cal H}(g')= \sum_{j=1}^n w_jg_{(j)}+t[w_{i_g}-w_{(i-1)_g}]. $$
But now, we know that $w_{i_f}-w_{(i-1)_f}=w_{i_g}-w_{(i-1)_g}$ by
hypothesis of equidistance, and thus $f'\succeq g'.$ Therefore,
{\bf A7} holds. $ \cqd $

Then, we have proved the following:

\begin{theorem}\label{car2addsym}
\cite{migr00} Let $\succeq $ be a binary relation on $\mathcal{F}\times \mathcal{F}$. The
following statements are equivalent:
\begin{enumerate}
\item
$\succeq $ satisfies {\bf A1}, {\bf A2}, {\bf A3}, {\bf A4}, {\bf
A5}, {\bf A6} and {\bf A7}.
\item
There is a unique 2-additive (or 1-additive) symmetric capacity $\mu $ such
that $\succeq $ is represented by ${\cal C}_{\mu }.$
\end{enumerate}
\end{theorem}

Remark that we have not used {\bf A8}. If we add {\bf A8}, it follows that the
weights are strictly decreasing:

\begin{lemma}\label{carbel}
Let $\succeq $ be a preference relation over $\mathcal{F}\times
\mathcal{F}$. If {\bf A1}, {\bf A2}, {\bf A3}, {\bf A4}, {\bf A5}, {\bf A6}
hold, then {\bf A8} is equivalent to $$w_1> w_2> \cdots > w_n.$$
\end{lemma}

{\bf Proof:} We already know that the aggregation operator satisfying {\bf A1},
{\bf A2}, {\bf A3}, {\bf A4}, {\bf A5}, {\bf A6} is an OWA operator (Theorem
\ref{carowa}). Then, it can be written as
$$ {\cal H}(f)=\sum_{i=1}^n f_{(i)} w_i,\, w_i\geq 0,\, \sum_{i=1}^n w_i=1.$$
Now, consider $f$ and $f'$ as in {\bf A8}. It is easy to see that $$ {\cal
H}(f)- {\cal H}(f')=w_{i+1}t-w_i t.$$ Suppose {\bf A8} holds. Then, it follows
that $w_i>w_{i+1}.$

Conversely, if $w_i>w_{i+1}$, {\bf A8} holds. $\cqd $

Then, if we add {\bf A8} in Theorem \ref{car2addsym}, we obtain a
characterization of 2-additive symmetric capacities with strictly
decreasing weights in the OWA operator. But this means $m(i,j)>0$
(as $m(i,j)=w_i-w_{i+1}$). On the other hand, by monotonicity
(Proposition \ref{monconmob}) we have $m(i)\geq 0.$ Hence, the
2-additive symmetric measure is also a belief function by
Proposition \ref{conbel}. We write down this result in next
corollary:

\begin{corollary}\label{carbel2addsym}
Let $\succeq $ be a binary relation on $\mathcal{F}\times \mathcal{F}$. The
following are equivalent:
\begin{enumerate}
\item
$\succeq $ satisfies {\bf A1}, {\bf A2}, {\bf A3}, {\bf A4}, {\bf
A5}, {\bf A6}, {\bf A7} and {\bf A8}.
\item
There is a unique 2-additive symmetric capacity $\mu $ such that $\succeq $ is
represented by ${\cal C}_{\mu }$ and the weights of the OWA operator are
strictly decreasing.
\item
There is a unique 2-additive symmetric belief function $\mu $ such that
$\succeq $ is represented by ${\cal C}_{\mu }.$
\end{enumerate}
\end{corollary}

In \cite{pogi94}, Theorem B, it is proved that {\bf A1}, {\bf A2}, {\bf A3}, {\bf A6}, {\bf A7} and {\bf A8} characterize the Gini index. Consequently, a preference relation modelled through the Choquet integral w.r.t. a 2-additive symmetric measure is a special case of the Gini index, just imposing order preserving gift and monotonicity ({\bf A4, A5}). 

Remark that for this last result, we avoid the possibility of
1-additive belief functions (i.e. probabilities), as $m(i,j)\not=
0.$ If we want to allow this possibility, {\bf A8} should be
changed into a weaker version:

\begin{itemize}
\item {\bf A8'.} Weak inequality aversion: For every $f, f'$ in
$\mathcal{F}_M$, for all $i\in X$, if $f'_i=f_i+t,
f'_{i+1}=f_{i+1}-t$ for some $t>0$, if $f'_j=f_j$ for all $j\not=
i, i+1$, then $f'\succeq f$.
\end{itemize}

\section{Characterization of $k$-additive OWA}

Let us now turn to the general $k$-additive symmetric case. From the precedent
analysis, we just have to replace {\bf A7} by another axiom for the
$k$-additive case, which we call {\bf A7(k)}. We follow the same sequence as in
the previous section.

\begin{proposition}\label{pesmobkadd}\cite{gra97b}
Let $\mu $ be a symmetric capacity, $m$ its M\"obius transform, and $w$ the
weight function of the corresponding OWA operator. If $\mu $ is at most
$k$-additive, then $$ \sum_{j=0}^{k-1}(-1)^j{k-1\choose j}w_{i+j}=c_k,\,
\forall i=1,\ldots, n-k+1,
$$ and moreover $c_k=m (A)$, with $|A|=k$.
\end{proposition}

The reciprocal result is given by:

\begin{proposition}
Let $w$ be the weight vector of an OWA operator. If $$
\sum_{j=0}^{k-1}(-1)^j{k-1\choose j}w_{i+j}= \sum_{j=0}^{k-1}(-1)^j{k-1\choose
j}w_{l+j},\quad i,l=1, \ldots, n-k+1,$$ then the corresponding capacity $\mu $
is at most $k$-additive.
\end{proposition}

{\bf Proof:} We need a preliminary result:

\begin{lemma}\label{aux3kadd}
Let $\mu $ be a symmetric capacity, $w$ the weight vector of the
corresponding OWA, and $A\subseteq X$. If $|A|\geq k$, then
$\forall \{ i_1,\ldots,i_k\} \subseteq A$, we have that $$
\sum_{B\subseteq \{ i_1,\ldots,i_k\} }\mu (A\setminus
B)(-1)^{|B|}=\sum_{j=0}^{k-1}(-1)^j{k-1\choose j}w_{n-|A|+j+1}.$$
\end{lemma}

{\bf Proof:} We know that the capacity associated to an OWA is
given by (see the proof of Lemma \ref{axisym}): $$ \mu
(T)=\sum_{i=n-|T|+1}^n w_i,\quad T\subseteq X,\quad T\not=
\emptyset .$$ Let us fix $j$ and compute how many times each $w_j$
appears. $w_j$ appears in any $\mu (T)$ such that $$ n-|T|+1\leq j
\Leftrightarrow |T|\geq n-j+1.$$ In our case, $T=A\setminus B,$
whence $w_j$ appears in any $B$ such that
\begin{equation}\label{auxlemk1}
|B|\leq |A|+j-n-1.
\end{equation}
We have three different cases:

\begin{itemize}
\item If $|A|+j-n-1<0,$ i.e. $j<n+1-|A|,$ then no $B$ satisfies
the conditions of (\ref{auxlemk1}). It follows that $w_j$ never
appears in the definition of $\mu (A\setminus B)$, whatever $B$
considered.

\item Our second case stands when $|A|+j-n-1\geq k,$ i.e. $j\in \{
n-|A|+k+1,\ldots,n\} $ (if $|A|=k$, there is no $j$ in these
conditions). Then, any $|B|$ satisfies (\ref{auxlemk1}), whence it
follows that $w_{j}$ appears in the definition of any
 $\mu (A\setminus B).$ Thus, $w_j$ appears
 $\displaystyle{\sum_{i=0}^k{k\choose i}(-1)^i=(1-1)^k=0}$ times.

\item Finally, assume $0\leq |A|+j-n-1< k,$ i.e. $j\in \{ n+1-|A|,
\ldots, n-|A|+k\} .$ In this case, $w_j$ appears in the definition
of all $A\setminus B$ satisfying such that $|B|\leq |A|+j-n-1:=l.$
For each $|B|$, we have ${k\choose |B|}$ subsets in these
conditions, whence it suffices to prove
$$\sum_{p=0}^{l}(-1)^p{k\choose p}=(-1)^{l}{k-1\choose l}, l\in \{
0,\ldots,k-1\} ,$$ but this is a well-known result in
Combinatorics (see e.g. \cite{ber71}). Then, the coefficient of
$w_j$ (with $j=n-|A|+1+l$) is $(-1)^{l}{k-1\choose l}$ and the
result holds. $\cqd $
\end{itemize}

Let us now prove the proposition. Suppose $\mu $ is not $k$-additive. Then, it
must exist $A$ such that $|A|>k$ and $m(A)\not= 0$. Let us choose such an $A$
of minimal cardinality.

Consider $\{ i_1, ..., i_k\} .$ Then, by Lemma \ref{aux3kadd}, we
know that $$ \sum_{B\subseteq \{ i_1,\ldots,i_k\} }\mu (i_1,
\ldots ,i_k\setminus
B)(-1)^{|B|}=\sum_{j=0}^{k-1}(-1)^j{k-1\choose j}w_{n-k+j+1}.$$
For a fixed $C\subseteq \{ i_1,\ldots,i_k\} $, let us find the
coefficient of $m(C)$ in $\sum_{B\subseteq \{ i_1,\ldots,i_k\}
}\mu (i_1,\ldots,i_k\setminus B)(-1)^{|B|}$ expressed with
the M\"obius transform through Eq. (\ref{eq:mob}). For this, we
have to count the number of times each $m(C)$ appears. Given
$C\subseteq \{ i_1, \ldots ,i_k\} ,$ the corresponding $m(C)$
appears in the definition of any $\mu (i_1, \ldots ,i_k\setminus B)$ such that $C\subseteq \{ i_1, \ldots ,i_k\}
\setminus B.$ For fixed $|B|$ we have ${k-|C|\choose |B|}$ subsets
in these conditions. Consequently, if $|C|<k,$ it follows that
$m(C)$ appears
$$ \sum_{|B|=0}^{k-|C|}{k-|C|\choose |B|}(-1)^{|B|}=0 $$ times. If
$C=\{ i_1,\ldots ,i_k\} ,$ then $m(C)$ appears once (when
$B=\emptyset ),$ whence $$ \sum_{B\subseteq \{ i_1,\ldots,i_k\}
}\mu (i_1, \ldots ,i_k\setminus B)(-1)^{|B|}= m(i_1, \ldots
, i_k).$$

Let us now prove that $\forall \{ i_1,\ldots,i_k\} \subset A$, we
have that
$$ \sum_{B\subseteq \{ i_1,\ldots,i_k\} }\mu (A\setminus
B)(-1)^{|B|}=m(A)+m(i_1\cdots i_k).$$

Let us consider $C\subseteq A$. Then, $m(C)$ appears in the
expression of $\mu (A\setminus B)$ in terms of the M\"obius
transform for all $B$ such that $C\subseteq A\setminus B$. Then,
it follows that $B\subseteq \{ i_1,\ldots, i_k\} \setminus C$. We
have the following cases:
\begin{itemize}
\item If $|\{ i_1,\ldots, i_k\} \setminus C|\not= 0$, the
coefficient of $m(C)$ is $$ \sum_{i=0}^{|\{ i_1,\ldots, i_k\}
\setminus C|}(-1)^i{|\{ i_1,\ldots, i_k\} \setminus C|\choose
i}=(1-1)^{k-|C\cap \{ i_1,\ldots,i_k\} |}=0.$$

\item When $|\{ i_1,\ldots, i_k\} \setminus C|=0$, the only
possible $B$ is $B=\emptyset $ and in this case the coefficient is
1.
\end{itemize}

On the other hand, if $|C|>k, C\not= A$, then $m(C)=0$ as $A$ is of minimal cardinality satisfying $m(A)\not= 0, |A|>k.$ Thus, the only possible non-null summands are $m(A)$
and $m(i_1,\ldots, i_k)$, both of them multiplied by 1. As a
conclusion, we obtain $$ \sum_{B\subseteq \{ i_1,\ldots,i_k\} }\mu
(A\setminus B)(-1)^{|B|}=m(A)+m(i_1\cdots i_k).$$

Therefore, $\displaystyle{\sum_{j=0}^{k-1}(-1)^j{k-1\choose
j}w_{n-|A|+j+1}}$ is not constant in $A$ and this contradicts our
hypothesis, whence the result holds. $ \cqd $

Let us now consider the following axiom \cite{migr00}:

\begin{itemize}
\item {\bf A7(k)} $k$-dimensional order preserving transfer:
$\forall f, g, f', g'\in \mathcal{F}$ and $i\in X $ such that
$i$ precedes $i+1$, which precedes $i+2$,\ldots, which precedes
$i+k$, for $f, f', g$ and $g'$, if
\begin{enumerate}
\item $f'$ and $g'$ are defined for some $t>0$ by
$$f'_{i+j}=f_{i+j}+(-1)^j{k-1\choose j}t, j=0,\ldots , k-1,
\, f'_k=f_k, k\not= i, i+1, \ldots, i+k-1.$$
$$g'_{i+j}=g_{i+j}+(-1)^j{k-1\choose j}t, j=0,\ldots , k-1,
\, g'_k=g_k, k\not= i, i+1, \ldots, i+k-1. $$

\item $f, f'$ and $g, g'$ are comonotone,
\end{enumerate}
then $f\succeq g\Leftrightarrow f'\succeq g'$.
\end{itemize}

Remark that this axiom, despite being a generalization of {\bf
A7}, is rather difficult to translate into natural language. We
will deal with the problem of interpretation at the end of this
section.

The following proposition holds:

\begin{proposition}\label{propo11}
Let $\succeq $ be a binary relation on $\mathcal{F}\times
\mathcal{F}$. If $\succeq $ satisfies {\bf A1}, {\bf A2}, {\bf
A3}, {\bf A4}, {\bf A5}, {\bf A6}, then {\bf A7(k)} is equivalent
to $$\sum_{j=0}^{k-1}(-1)^j{k-1\choose
j}w_{i+j}=\sum_{j=0}^{k-1}(-1)^j{k-1\choose j}w_{i'+j},\, \forall
i,i'=1,\ldots, n-k+1,$$ i.e. $
\displaystyle{\sum_{j=0}^{k-1}(-1)^j{k-1\choose j}w_{i+j}}$ does
not depend on $i$.
\end{proposition}

{\bf Proof:} $\Rightarrow )$ The proof is an adaptation of the one
of Lemma 3.2.1 in \cite{pogi94}:

First, it will prove useful to focus on the interior of $\mathcal{F}_M$, denoted $(\mathcal{F}_M)^o$ and given by 
$$(\mathcal{F}_M)^o=\{ f\in \mathcal{F}_M\, |\, f_1<f_2< \cdots < f_n\} .$$

Let $f\in (\mathcal{F}_M)^o$ and take $\epsilon >0$. Consider the
auxiliary acts $e^i$ defined by $e^i(j)=1$ if $j=i$ and $e^i(j)=0$
otherwise. We define
$$f'=f+\sum_{j=0}^{k-1}\epsilon e^{i+j}(-1)^{j-1}{k-1\choose j-1}.$$

Here $\epsilon>0 $ is small enough to maintain the order. Remark
that we can take $\epsilon>0 $ because $f\in (\mathcal{F}_M)^o$,
i.e. we are in the set of strictly increasing acts. Now, for fixed $i$, let
$\pi $ be the permutation on $X$ defined by

\begin{enumerate}
\item If $i>k,\, \pi (1)=i, \pi (2)=i+1, \ldots, \pi (k)=i+k-1,
\pi (i)=1, \pi (i+1)=2, \ldots,$

$$ \pi (i+k-1)=k, \pi (l)=l, l\notin \{1,\ldots,k,i,\ldots,i+k-1\} .$$
\item If $i\leq k,\, \pi (1)=k+1, \pi (2)=k+2, \ldots, \pi
(i-1)=i+k, \pi (i)=1, \pi (i+1)=2, \ldots,$

$$\pi (i+k-1)=k, \pi (l)=l, l\notin \{ 1,\ldots,i+k-1\} .$$
\end{enumerate}
Consider $g=\pi f.$ Then, $g'$ in the conditions of {\bf A7(k)} is
given by
$$g'=g+\sum_{j=1}^k\epsilon e^j(-1)^{j-1}{k-1\choose j-1}.$$

By symmetry {\bf A3}, we have $g\sim f.$ Then, by {\bf A7(k)}, we
get $f'\sim g'$ and thus $$ {\cal H}(f')-{\cal H}(f)={\cal
H}(g')-{\cal H}(g).$$ Consequently, as ${\cal H}$ is indeed an OWA
operator,
$$ {\cal H}(f')-{\cal H}(f)=\sum_{j=1}^k\epsilon w_j(-1)^{j-1}{k-1\choose
j-1}=\sum_{j=1}^k\epsilon w_{i+j-1}(-1)^{j-1}{k-1\choose
j-1}={\cal H}(g')-{\cal H}(g),$$ whence the result.

$\Leftarrow )$ Consider $f, g, f', g'\in \mathcal{F}$ in the conditions
of {\bf A7(k)}, and let us suppose that $f\succeq g$. Then, we
have $$ {\cal H}(f')=\sum_{i=1}^nf_{(i)}w_i+t\left[
\sum_{j=0}^{k-1}(-1)^j{k-1\choose j}w_{i+j}\right] .$$

$$ {\cal H}(g')=\sum_{i=1}^ng_{(i)}w_i+t\left[
\sum_{j=0}^{k-1}(-1)^j{k-1\choose j}w_{i'+j}\right] .$$ By
hypothesis, we know that $$\sum_{j=0}^{k-1}(-1)^j{k-1\choose
j}w_{i+j}=\sum_{j=0}^{k-1}(-1)^j{k-1\choose j}w_{i'+j}.$$ Then, we
obtain that $f'\succeq g'$ and {\bf A7(k)} holds. $\cqd $

Summarizing, we have proved the following:

\begin{theorem}\label{carowaksym}
\cite{migr00} Let $\succeq $ be a binary relation on $\mathcal{F}\times
\mathcal{F}$. The following are equivalent:
\begin{enumerate}
\item
$\succeq $ satisfies {\bf A1}, {\bf A2}, {\bf A3}, {\bf A4}, {\bf
A5}, {\bf A6} and {\bf A7(k)}.
\item
There is a unique at most $k$-additive symmetric capacity
$\mu $ such that $\succeq $ is represented by ${\cal C}_{\mu }.$
\end{enumerate}
\end{theorem}

Let us come back to the problem of interpreting {\bf A7(k)}. The
fact that this axiom is much more difficult to deal with than {\bf
A7} constitutes a weakness in this characterization. A possible
interpretation has been given by T. Gajdos in \cite{gad02}, using an equivalent
axiom which reads:
\begin{itemize}
\item {\bf A7'(k).}  $\forall f\in \mathcal{F}_M, \forall i,
j\in \{ 1,\ldots , n-k+1\} , \forall t>0$,
\begin{enumerate}
\item
Define $f^i$ and $f^j$ by
$$f^i_{i+r}=f_{i+r}+(-1)^r{k-1\choose r}t, r=0,\ldots , k-1,\, f^i_l=f_l,
l\not= i, i+1, \ldots, i+k-1.$$
$$f^j_{j+r}=f_{j+r}+(-1)^r{k-1\choose r}t, r=0,\ldots , k-1,\, f^j_l=f_l,
l\not= j, j+1, \ldots, j+k-1. $$ \item If $f, f^i, f^j$ are
comonotone,
\end{enumerate}
then $f^i\sim f^j$.
\end{itemize}

\begin{proposition}
Consider a preference relation over $\mathcal{F}\times
\mathcal{F}$. If {\bf A1}, {\bf A2}, {\bf A3}, {\bf A4}, {\bf
A5}, {\bf A6} hold, then {\bf A7'(k)} is equivalent to
$$\sum_{j=0}^{k-1}(-1)^j{k-1\choose
j}w_{i+j}=\sum_{j=0}^{k-1}(-1)^j{k-1\choose j}w_{i'+j},\, \forall
i,i'=1,\ldots, n-k+1.$$
\end{proposition}

{\bf Proof:} If {\bf A1}, {\bf A2}, {\bf A3}, {\bf A4}, {\bf A5},
{\bf A6} hold, we already know (Theorem \ref{carowa}) that the
functional ${\cal H}$ representing $\succeq $ is the Choquet
integral with respect to a symmetric capacity $\mu .$ Let us
denote by $w$ the corresponding weight vector. Then, $$ {\cal
H}(f)= \sum_{i=1}^n f_{(i)} w_i, \, \forall f\in {\cal F}.$$

Consider $f, f^i, f^j$ fulfilling {\bf A7'(k)}. It follows that
$$ {\cal H}(f^i)= {\cal H}(f)+ \sum_{r=0}^{k-1}(-1)^r t
{k-1\choose r}w_{i+r}, \, {\cal H}(f^j)={\cal H}(f)+
\sum_{r=0}^{k-1}(-1)^r t {k-1\choose r}w_{j+r}.$$ Then, $f^i\sim
f^j$ if and only if
$$\sum_{j=0}^{k-1}(-1)^j{k-1\choose
j}w_{i+j}=\sum_{j=0}^{k-1}(-1)^j{k-1\choose j}w_{i'+j},\, \forall
i,i'=1,\ldots, n-k+1,$$ whence the result. $\cqd $

Then, by Proposition \ref{propo11}, we can interchange {\bf A7(k)}
and {\bf A7'(k)} if {\bf A1}, {\bf A2}, {\bf A3}, {\bf A4}, {\bf
A5}, {\bf A6} hold. In \cite{gad02}, Gajdos also showed the equivalence of these
two axioms. 

In {\bf A7'(k)} the quantities added to individuals can be
interpreted as a reward scheme. With this interpretation in mind,
let us consider first a reward of $\epsilon $ to an individual.
Now, we ask the following question: Does the decision maker cares
about which individual receives this reward? If not, then the
decision maker does not care about equality as for him it is the
same to give a reward to the richest individual as to give a
reward to the poorest one. But if the answer is yes, the decision
maker cares about inequality. Since $f$ is strictly increasing, we
may think that if the decision maker cares about inequality, then
$f^i\succeq f^{i+1}$, or equivalently, $$ (f_1, \ldots,
f_i+\epsilon ,f_{i+1}, \ldots, f_n)\succeq (f_1,\ldots, f_i,
f_{i+1}+\epsilon , \ldots, f_n)$$ which is in turn equivalent to
$$ (f_1, \ldots, f_i+\epsilon ,f_{i+1}-\epsilon , \ldots,
f_n)\succeq (f_1,\ldots, f_i, f_{i+1} , \ldots, f_n),$$ provided
the order in the incomes has not changed. But now, we can go a
step further and ask the decision maker if he cares about which
individual $i$ is considered in this new reward scheme. If he does
not care, then the decision maker satisfies {\bf A7'} (or {\bf
A7'(2)}). In other case, he is quite sensitive to inequalities and
we may think that
$$ (f_1, \ldots, f_i+\epsilon ,f_{i+1}-\epsilon , f_{i+2} \ldots,
f_n)\succeq (f_1,\ldots, f_i, f_{i+1}+\epsilon , f_{i+2}-\epsilon
, \ldots, f_n),$$ or in other words, it is
$$ (f_1, \ldots, f_i+\epsilon ,f_{i+1}-2\epsilon , f_{i+2}+\epsilon ,
\ldots, f_n)\succeq (f_1,\ldots, f_i, f_{i+1} , f_{i+2}, \ldots,
f_n),$$ provided the order in the incomes has not changed. We can
repeat the process for this scheme of rewards to sharpen the
degree in which the decision maker cares about inequalities. Then,
{\bf A7'(k)} for different choices of $k$ can be seen as a scale to
measure the sensitivity to inequalities of the decision maker.

Another possibility of characterization for $k$-additive symmetric
measures could be derived from the results of Calvo and De Baets
\cite{caba98} and Cao-Van and De Baets \cite{caba01}, in which
they introduce the concept of binomial OWA operators.

\begin{definition}
Let $k\in \{ 1,\ldots, n\} $. The {\bf $k$-binomial OWA operator}
is the OWA operator with weight vector $w:=(w_{k1},\ldots,
w_{kn})$ defined by $$ w_{ki}={{n-i\choose k-1}\over {n\choose
k}}.$$
\end{definition}

Then, they prove the following:

\begin{theorem}
Let $k\in \{ 1,\ldots, n\} $. Consider an aggregation operator
${\cal H}$; then, the following equivalence holds:
\begin{enumerate}
\item ${\cal H}$ is the Choquet integral with respect to a
symmetric $k$-additive capacity on $X$. \item ${\cal H}$ is a
weighted sum of the first $k$-binomial OWA operators.
\end{enumerate}
\end{theorem}

This result characterizes $k$-additive symmetric capacities from a
mathematical point of view. We feel that this result can help to
characterize $k$-additive symmetric measures. However, a wide
study of the binomial OWA operators must be done. Besides, we need
an axiom leading to a representation in terms of a weighted sum.

We have to remark as a conclusion that the properties that
characterize the $k$-additive symmetric case are rather special.
For the general $k$-additive case, it can be seen that these
properties do not hold. In fact, the most important tool that we
have used in the proofs of the results was the fact that
$m(A)=m(B)$ whenever $|A|=|B|$, and this property comes from the
symmetry. This implies that even if these results can give us some
information about the general case, this one is much more
difficult to handle.

This will not be the case for general measures. We will prove in the
next section that removing the symmetry axiom, we can obtain a
characterization of Choquet integral.

\section{Characterization of Choquet integral for the general case}

>From an interpretational point of view, the difference between the
symmetric case and the general case is given by the fact that for
the symmetric case, the importance of an individual when computing
Choquet integral comes from its relative position, while for the
general case its importance depends also on its index.

To clarify this idea, suppose that we are in the symmetric case
and we have an income distribution $f$. Then, if we give a reward
of $t$ to an individual $i$ so that the order in the income
distribution does not change, the amount of welfare is determined
by the relative position of the income of $i$, but we do not
really care about which individual is considered. This is not true
in general for the non-symmetric situation. If we are in the
general case, a gift of $t$ to individual $i$ determines an
increment of welfare depending on the relative position of
individual $i$ and the concrete individual who has received the
gift.

To characterize such preference relations, it is obvious that the
symmetry axiom {\bf A3} must be removed. In order to prove our
result we need the following result due to Schmeidler in which we
obtain the mathematical conditions that characterize Choquet
integral:

\begin{theorem}\cite{sch86}\label{thesch}
Let $X$ be a universal set (finite or infinite) and $ {\cal X}$ be
a $\sigma $-algebra over $X$. Consider ${\cal B}(X, {\cal X} )$
(or ${\cal B}$ for short) the set of bounded, real valued, $\sigma
$-measurable functions on $X$ (i.e. the set of random variables).
Let ${\cal H}:{\cal B}\rightarrow \mathbb{R}$ satisfying ${\cal
H}(1_X)=1$ be given. Suppose also that the functional ${\cal H}$
satisfies
\begin{enumerate}
\item Comonotonic additivity: $f$ and $g$ comonotonic implies
${\cal H}(f+g)={\cal H}(f)+{\cal H}(g).$ \item Monotonicity:
$f(x)\geq g(x),\, \forall x\in X$ implies ${\cal H}(f)\geq {\cal
H}(g).$
\end{enumerate}
Then, defining $\mu (A)={\cal H}(1_A),\, \forall A\in {\cal X}$ we
have $$ {\cal H}(f)=\int_0^{\infty }\mu (f\geq \alpha )d\alpha +
\int_{-\infty }^0 (\mu (f\geq \alpha )-1)d\alpha .$$
\end{theorem}

This expression is the Choquet integral for real functions
(Definition \ref{cho}). In our case, in which we are treating with
finite universal sets, we can consider ${\cal X} ={\cal P}(X)$ and
thus any act is measurable; and as $X$ is finite, any act is also
bounded.

Now, the following can be proved:

\begin{theorem}\label{carcho}
\cite{migr00} Let $\succeq $ be a binary relation on $\mathcal{F}\times
\mathcal{F}$. The following statements are equivalent:
\begin{enumerate}
\item
$\succeq $ satisfies {\bf A1}, {\bf A2}, {\bf A4}, {\bf A5} and
{\bf A6}.
\item
There is a unique capacity $\mu $ such that
$\succeq $ is represented by ${\cal C}_{\mu }.$
\end{enumerate}
\end{theorem}

{\bf Proof:} It is clear that Choquet integral satisfies all these
axioms. Let us see that this is the only functional for which
these conditions hold.

Thus, let ${\cal H}$ be a functional satisfying {\bf A1}, {\bf
A2}, {\bf A4}, {\bf A5}, {\bf A6}. We will denote by ${\cal
H}_{\pi }$ the restriction of ${\cal H}$ to a simplex $H_{\pi },$
which is defined as follows: for a given permutation $\pi $ of the
indices, $H_{\pi }=\{ (x_1,\ldots,x_n)\, |\, x_{\pi (1)}\leq
x_{\pi (2)}\leq \ldots\leq x_{\pi (n)}\} $. Now, we extend ${\cal
H}_{\pi }$ to $\mathbb{R}^n_+$ by symmetry, i.e. given
$(x_1,\ldots,x_n),$ we take a permutation $\alpha $ such that
$(x_{\alpha (1)},\ldots ,x_{ \alpha (n)})\in H_{\pi }$ and we
define $\bar{\cal H}(x_1,\ldots,x_n)={\cal H}_{\pi }(x_{\alpha
(1)},\ldots , x_{\alpha (n)}).$ Thus, $\bar{\cal H}$ is a
symmetric functional. Symmetry axiom, together with {\bf A1}, {\bf
A2} and {\bf A4} is the set of axioms used by Weymark in
\cite{wey81} (Theorem \ref{wey}). Then, we know that $\bar{\cal
H}$ is given by
$$ \bar{\cal H}(f)=\sum_{i=1}^np_if_{(i)},$$ for some $ p_i\in
\mathbb{R},\, i=1,\ldots, n,$ and therefore $\bar{\cal H}$
satisfies comonotonic additivity.

Then, $\bar{\cal H}$ is linear in each simplex. In particular
$(\bar{\cal H})_{\pi }$ is linear. But $(\bar{\cal H})_{\pi
}={\cal H}_{\pi }$. Since this is valid for any $\pi $, ${\cal H}$
itself is linear in each simplex.

Now, we define $ \mu
(A)={\cal H}(1_A).$

By {\bf A5} and {\bf A6} we know that $\mu $ is a capacity. Then,
we have the conditions of the result of Schmeidler (Theorem
\ref{thesch}) characterizing the Choquet integral. So that
$\succeq $ is represented by a Choquet integral and thus the
result holds. $ \cqd $

Remark that this result is just a version of the conditions of
Schmeidler while having a preference relation over the set of
acts. In fact, {\bf A4} is a version for acts of the comonotonic
additivity.

This result is very similar to the one proved by Chateauneuf in
\cite{cha94}.

He uses the following set of axioms:

\begin{itemize}
\item {\bf B1} $\succeq $ is a non-trivial weak order.

\item {\bf B2} Continuity with respect to monotone uniform
convergence
\begin{enumerate}
\item $f_n, f, g\in {\cal F}, f_n\succeq g, f_n\downarrow ^u
f\Rightarrow f\succeq g.$

\item $f_n, f, g\in {\cal F}, g\succeq f_n, f_n\uparrow ^u
f\Rightarrow g\succeq f.$
\end{enumerate}

\item {\bf B3} Monotonicity $f_i\geq g_i+\epsilon ,\forall i$
(where $\epsilon
>0$ is a constant) $\Rightarrow f\succ g.$

\item {\bf B4} Comonotonic independence $f, g, h\in {\cal F}, f$
and $h$ comonotonic, $g$ and $h$ comonotonic, then $f\sim
g\Rightarrow f+h\sim g+h.$
\end{itemize}

and then he proves the following:

\begin{theorem}\label{cha}
Let $\succeq $ be a binary relation on ${\cal F}\times {\cal
F}$. The following are equivalent:
\begin{enumerate}
\item $\succeq $ satisfies {\bf B1}, {\bf B2}, {\bf B3} and {\bf
B4}.

\item there is a unique capacity $\mu $ such that $\succeq $ is
represented by ${\cal C}_{\mu }.$
\end{enumerate}
\end{theorem}

It is easy to see that {\bf B1} is equivalent to {\bf A1} and {\bf
A6}, {\bf B2} is equivalent to {\bf A2}, {\bf B3} is similar to
{\bf A5} and finally {\bf B4} is implied by {\bf A4}. However,
Theorem \ref{cha} is not restricted to finite sets and he only imposes $f$ to be a bounded measurable function.

\section{Characterization of 2-additive capacities}

Now, we deal with the problem of characterizing preference
relations on $\mathcal{F}\times \mathcal{F}$ that can be
represented by the Choquet integral with respect to a 2-additive
capacity. Of course, we only need to add an axiom to the set of
axioms found in last section for the general case. However, {\bf
A7} does not suffice to characterize 2-additivity and therefore
our new axiom must be a generalization of it.

Then, we have to find a new property characterizing 2-additive
capacities. This property is obtained in next proposition.

\begin{proposition}\label{car2add}
A capacity $\mu $ is at most 2-additive if and only if $ \forall
A\subseteq X,$ and $ \forall i,j\in A, $ $$ \mu (A)-\mu
(A\setminus i)-\mu (A\setminus j)+\mu (A\setminus i,j) =\mu (i,j)-\mu (i)-\mu (j).
$$
\end{proposition}

{\bf Proof:} Let us suppose that $\mu $ is not a 2-additive
capacity (at most). Then, there exists $A$ such that $|A|>2$ and
$m(A)\not= 0$.

Let us consider $A$ of minimal cardinality in these conditions and
let us take $i,j\in A$. This is always possible as $|A|>2$. Then,
we have the following: $$ \mu (A)-\mu (A\setminus i)-\mu
(A\setminus j)+ \mu (A\setminus ij)= \sum_{B\subseteq
A}m(B)-\sum_{B\subseteq A\setminus i}m(B)-\sum_{B\subseteq
A\setminus j}m(B)+ \sum_{B\subseteq A\setminus ij}m(B)$$
$$=\sum_{i\in B\subseteq A}m(B)- \sum_{i\in B\subseteq A\setminus
j}m(B)=m(A)+m(ij), $$ by hypothesis on $A$.

On the other hand, $ \mu (ij)-\mu (i)-\mu (j)=m(ij) $ and thus the
expression does not remain constant.

Reciprocally, proceeding the same way, we obtain that if
$m(A)=0,\quad \forall A$ s.t. $|A|>2$, then $$ \mu (A)-\mu
(A\setminus i)-\mu (A\setminus j)+\mu (A\setminus ij)= m(ij)=\mu
(ij)-\mu (i)-\mu (j), $$ whence the result. $\cqd $

Let us introduce the following notation: Given an act $f\in
\mathcal{F}$, for a fixed real value $t$, we define the act
$f^B\in \mathcal{F}$ by
\[ f^B_i=\left\{
\begin{array}{ll} f_i+t & {\rm if }\, i\in B\\ 0 & {\rm
otherwise}\end{array}\right. \]

Consider now the following axiom:

\begin{itemize}
\item {\bf A9.} Let $f, g\in \mathcal{F}$ be acts such that
$f_{i}=f_{j}, g_{i}=g_{j}.$ Consider the acts $f^B, g^B$ for all
$B\subseteq \{ i, j\} ,$ with $t>0$ so that $f^B, f$ and $g^B, g$ are comonotone for all $B.$ Let ${\cal H}:\mathcal{F}\rightarrow
\mathbb{R}$ be any functional representing the preference relation
$\succeq $. Then,
$$\sum_{B\subseteq \{ i, j\} }{\cal H}(f^B)
(-1)^{|B|}=\sum_{B\subseteq \{ i, j\} }{\cal H}(g^B)(-1)^{|B|}.$$
\end{itemize}

This axiom can be interpreted as follows: Suppose that we have an
income distribution such that two individuals $i, j$ are equally
rich. Now, we give a gift $t>0$ to both of them. Next step is to
ask one of these individuals for this gift to be returned. This
individual can be either $i$ or $j$. Then, we obtain four income
distributions and the global welfare with these operations could
have changed. If the amount (or decrease) in the welfare just
depend on individuals $i, j$ but not on their relative position,
then the decision maker acts following {\bf A9}.

Let us introduce the following notation: For any $h\in
\mathcal{F}$ we denote by $A^i_h$ the set $$A^i_h:=\{ l\in X\,
|\, h_l>h_i\} .$$

The following can be proved:

\begin{proposition}
Let $\succeq $ be a preference relation on $\mathcal{F}\times
\mathcal{F}$. If $\succeq $ satisfies {\bf A1}, {\bf A2}, {\bf
A4}, {\bf A5} and {\bf A6}, then {\bf A9} is equivalent to $$ \mu
(A)-\mu (A\setminus i)-\mu (A\setminus j)+\mu (A\setminus i,j)=\mu
(i,j)-\mu (i)-\mu (j),\quad \forall A\subseteq X,\quad i,j\in A.$$
\end{proposition}

{\bf Proof:} As {\bf A1}, {\bf A2}, {\bf A4}, {\bf A5} and {\bf
A6} hold, we know from Theorem \ref{carcho} that our preference
relation is modelled by the Choquet integral with respect to a
capacity $\mu $.

Consider $f, g\in \mathcal{F}$ fulfilling the conditions of {\bf
A9}. Then, it can be easily seen that
$$ {\cal C}_{\mu }(f^{\{ i,j\} })= {\cal C}_{\mu }(f)+t[\mu (A^i_f\cup i,j)-\mu
(A^i_f)].$$
 $$ {\cal C}_{\mu }(f^{\{ i\} })= {\cal C}_{\mu }(f)+t[\mu (A^i_f\cup i)-
 \mu (A^i_f)].$$
 $$ {\cal C}_{\mu }(f^{\{ j\} })= {\cal C}_{\mu }(f)+t[\mu (A_f^i\cup j)-
 \mu (A^i_f)].$$
The same can be done for $g$. Consequently, $$ \sum_{B\subseteq \{
i, j\} }{\cal C}_{\mu }(f^B)(-1)^{|B|}=(\mu (A^i_f\cup i,j)-\mu
(A^i_f\cup i)-\mu (A^i_f\cup j)+\mu (A^i_f))t,$$
$$ \sum_{B\subseteq \{ i, j\}
}{\cal C}_{\mu }(g^B)(-1)^{|B|}=(\mu (A^i_g\cup i,j)-\mu (A^i_g\cup
i)-\mu (A^i_g\cup j)+\mu (A^i_g))t,$$ whence {\bf A9} holds for any $f, g$ if and
only if
$$\mu (A^i_f\cup i,j)-\mu (A^i_f\cup i)-\mu (A^i_f\cup j)+
\mu (A^i_f)=\mu (A^i_g\cup ij)-\mu (A^i_g\cup i)-\mu (A^i_g\cup
j)+\mu (A^i_g),$$ whence the result. $\cqd $

Then, we have proved:

\begin{theorem}\label{ter2add}
Let $\succeq $ be a binary relation on $\mathcal{F}\times
\mathcal{F}$. The following are equivalent:
\begin{enumerate}
\item $\succeq $ satisfies {\bf A1}, {\bf A2}, {\bf A4}, {\bf A5},
{\bf A6} and {\bf A9}.

\item There is a unique 2-additive capacity $\mu $ such that
$\succeq $ is represented by ${\cal C}_{\mu }.$
\end{enumerate}
\end{theorem}

\section{Characterization of $k$-additive capacities}

In this section we deal with the general $k$-additive case. We
will follow the same line as for the symmetric case. Then, we will
use the results characterizing preference relations modelled
through Choquet integral with respect to a capacity and change
{\bf A9} for another axiom {\bf A9(k)} for the general
$k$-additive case.

For the $k$-additive case, we have the following result:

\begin{proposition}
A capacity $\mu $ is (at most) $k$-additive if and only if
$$\sum_{\stackrel{B\subseteq \{ i_1,\ldots,i_k\} }
{i_1,\ldots,i_k\in A}}\mu (A\setminus B)(-1)^{|B|}
=\sum_{B\subseteq \{ i_1,\ldots,i_k\} }\mu (i_1,\ldots, i_k\setminus B)(-1)^{|B|} , \forall A\, \mbox{ such that }
i_1,\ldots,i_k\in A.$$
\end{proposition}

{\bf Proof:} Suppose that $\mu $ is not $k$-additive. Then, there
exist $A$ such that $|A|>k$ and $m(A)\not= 0$. Let us consider $A$
of minimal cardinality in these conditions. Let us prove that
$\forall \{ i_1,\ldots,i_k\} \subseteq A$, we have
$$\sum_{B\subseteq \{ i_1,\ldots,i_k\} }\mu (A\setminus
B)(-1)^{|B|}=m(A)+m(\{ i_1, \ldots , i_k\} ).$$ Consider $C\subseteq A$ such
that $|C|=j$. We have the following cases:

\begin{itemize}
\item If $\{ i_1,\ldots,i_k\} \not\subseteq C$, then $m(C)$
appears $(1-1)^{k-|C\cap \{ i_1,\ldots,i_k\} |}=0$ times.

\item If $|C|>k, C\not= A$, then $m(C)=0$ by hypothesis and the
coefficient is not important.

\item If $C=A$, then $m(A)$ appears once (when $B=\emptyset $).

\item Finally, if $C= \{ i_1,\ldots,i_k\} $, then $m(C)$ appears
once (only in $\mu (A)$).
\end{itemize}

On the other hand, repeating the process for $\{ i_1,\ldots, i_k\}
,$ we have $$ \sum_{B\subseteq \{ i_1,\ldots,i_k\} }\mu (i_1,\ldots,i_k \setminus B)(-1)^{|B|}=m(i_1,\ldots, i_k).$$
Thus, the expression is constant if and only if $\mu $ is a
$k$-additive measure. $ \cqd $

We consider the axiom:

\begin{itemize}
\item {\bf A9(k)} $k$-dimensional asymmetric order-preserving
transfer: Consider $f, g\in \mathcal{F}$ such that
$f_{i_1}=\cdots=f_{i_k}, g_{i_1}=\cdots=g_{i_k}.$ We consider
$f^B, g^B$ for all $B\subseteq \{ i_1,\ldots,i_k\} ,$ with $t>0$
so that $f^B, f$ and $g^B, g$ are comonotone for all $B$. Let
${\cal H}:\mathcal{F}\rightarrow \mathbb{R}$ be any functional
representing the preference relation $\succeq $. Then,
$$\sum_{B\subseteq \{ i_1,\ldots,i_k\} }{\cal H}(f^B)
(-1)^{|B|}=\sum_{B\subseteq \{ i_1,\ldots,i_k\} }{\cal
H}(g^B)(-1)^{|B|}.$$
\end{itemize}

Axiom {\bf A9(k)} could be interpreted from {\bf A9} the same way
as {\bf A7'(k)} is interpreted from {\bf A7'}. Starting with the
conditions of {\bf A9}, if the decision maker does not care about
the relative position of $i, j,$ then he follows {\bf A9};
otherwise he cares about the position, and then we turn to {\bf
A9(3)} and we restart again. The process finishes when we find the
value of $k$. Then, {\bf A9(k)} measures again the sensitivity of
the decision maker to the relative positions of individuals.

The following can be proved:

\begin{proposition}
Let $\succeq $ be a preference relation on $\mathcal{F}\times
\mathcal{F}$. If $\succeq $ verifies {\bf A1}, {\bf A2}, {\bf
A4}, {\bf A5} and {\bf A6}, then  {\bf A9(k)} is equivalent to $$
\sum_{\stackrel{B\subseteq \{ i_1,\ldots,i_k\} }
{i_1,\ldots,i_k\in C}}\mu (C\setminus
B)(-1)^{|B|}=\sum_{B\subseteq \{ i_1,\ldots,i_k\} }\mu
(B)(-1)^{k-|B|}, \forall C\, {\mbox {\it such that }}
i_1,\ldots,i_k\in C.$$
\end{proposition}

{\bf Proof:} As  {\bf A1}, {\bf A2}, {\bf A4}, {\bf A5} and {\bf
A6} hold, we know from Theorem \ref{carcho} that our preference
relation is modelled by the Choquet integral with respect to a
capacity $\mu $.

Let us take $f, g\in \mathcal{F}$ in the conditions of {\bf
A9(k)}. Given $i_1, \ldots, i_k,$ consider $A^{i_k}_f$ (resp. $
A^{i_k}_g$). Then,
$$ {\cal C}_{\mu }(f_B)={\cal C}_{\mu }(f)+t\left[ \mu (B\cup
A^{i_k}_f)-\mu (A^{i_k}_f)\right] ,\, {\cal C}_{\mu }(g_B)={\cal
C}_{\mu }(g)+t\left[ \mu (B\cup A^{i_k}_g)-\mu (A^{i_k}_g)\right]
.$$ Now,
$$ \sum_{B\subseteq \{ i_1, ..., i_k\} }{\cal H}(f^B)(-1)^{|B|}=t\sum_{B\subseteq \{ i_1,\ldots, i_k\} }\mu
(B\cup A^{i_k}_f)(-1)^{|B|},$$ 

$$\sum_{B\subseteq \{ i_1, ..., i_k\} }{\cal H}(g^B)(-1)^{|B|}=t\sum_{B\subseteq \{ i_1,\ldots, i_k\} }\mu (B\cup
A^{i_k}_g)(-1)^{|B|},$$ and thus, {\bf A9(k)} holds if and only if
$$\sum_{B\subseteq \{ i_1,\ldots, i_k\} }\mu (B\cup A^{i_k}_f)
(-1)^{|B|}=\sum_{B\subseteq \{ i_1,\ldots, i_k\} }\mu (B\cup
A^{i_k}_g)(-1)^{|A|},\, \forall f, g,$$ whence the result holds. $\cqd $

Then, we have proved:

\begin{theorem}
Let $\succeq $ be a binary relation on $\mathcal{F}\times
\mathcal{F}$. The following statements are equivalent:
\begin{enumerate}
\item $\succeq $ satisfies {\bf A1}, {\bf A2}, {\bf A4}, {\bf A5},
{\bf A6} and {\bf A9(k)}. \item There is a unique $k$-additive
capacity $\mu $ such that $\succeq $ is represented by ${\cal
C}_{\mu }.$
\end{enumerate}
\end{theorem}

\section{Conclusions}

We have proposed a  characterization of preference relations
represented by the Choquet integral with respect to different
types of capacities, from symmetric capacities to general
$k$-additive measures, with an emphasis on 2-additive capacities.

Axioms for 2-additive measures (in the symmetric and in the
general case) can be interpreted from an economical point of view
and the theory of social welfare. Axioms for the $k$-additive case
are just a generalization of those from the 2-additive case. For
the symmetric case, an interesting interpretation of the axioms
was given by Gajdos, but to our knowledge, no interpretation has
been given for the general case. For this case an additional
problem arises, as the relative strength of each individual is not
the same. Then, we have to work with families of acts or income
distributions, that make the resulting axioms rather difficult to
handle and to translate into natural language (axioms ${\bf
A9(k)}$).

A drawback considering these last axioms is that the aggregation
function $\mathcal{H}$ intervenes into them. This could be
overcome at the price of introducing \emph{difference measurement}
\cite{krlusutv71} (instead of relying on ordinal measurement),
requiring from the decision maker to express a certain kind of
intensity of preferences.

Another possible way to find a characterization would be to
consider the Shapley interaction index \cite{gra96c} instead of
the M\"obius transform; however, the Choquet integral in terms of
Shapley interaction is rather complicated \cite{gra97d}, and thus
a deep study should be done.

\newpage

\bibliographystyle{alpha}

\bibliography{/home/pedro/Biblio/ae.bib,/home/pedro/Biblio/ot.bib,/home/pedro/Biblio/uz.bib,/home/pedro/Biblio/kn.bib,/home/pedro/Biblio/fj.bib}

\end{document}